# Saturated absorption spectroscopy and frequency locking of DBR laser on the D$_2$ transition of rubidium atoms

Davood Razzaghi[*][1,2], Ali MotazediFard[2], Marzieh Akbari[1], Seyed Ahmad Madani[2], Masoud Yousefi[2], Ali Allahi[2], Ghazal Mehrabanpajooh[2], Mohsen Shokrolahi[2], Hamid Asgari[2], and Zafar Riazi[2]

[1]*Photonics and Quantum Technologies Research School, Nuclear Science and Technology Research Institute (NSTRI), North Amirabad, Tehran, Iran*
[2]*Iran Center of Quantum Technologies (ICQT), Tehran, 15998-14713, Iran*
*e-mail: davrazzaghi@yahoo.com

Recent advancements in laser cooling and spectroscopy techniques have led to the generation of new sensors. These sensors are based on either cold or hot atoms, contributing to the field of atom-based quantum metrology. This field encompasses magnetometry, navigation, atomic interferometry, and secondary atomic clocks. In many modern physics experiments, the precise control of certain optical characteristics of light, such as its frequency, is crucial. For instance, the cooling process of a rubidium atom typically requires two laser frequencies that are accurately locked to specific atomic transitions. In this setup, a tunable laser, often referred to as the pump laser, is locked onto the atomic transition $5^2S_{1/2}\ F = 2 \to 5^2P_{3/2}\ F' = 2,3$ (This is a common choice for the pump laser). Simultaneously, a second laser, known as the repump laser, is locked onto a different atomic transition, specifically $5^2S_{1/2}\ F = 1 \to 5^2P_{3/2}\ F' = 1,2$. The frequency difference between these two transitions, which corresponds to the hyperfine splitting of the atomic levels, is approximately 6.5 GHz. (Figure 1) [1 and 2].

The laser cooling process is highly sensitive to the frequency and shift between the pump laser and the desired energy level. Therefore, it's crucial to stabilize the laser frequency, ensuring it remains constant against rapid and slow changes over an extended period. This is known as the frequency locking technique. Various methods are employed to stabilize the laser frequency, among these methods, we can mention the Dichroic Atomic Vapor Laser Lock method, DAVLL, which utilizes a magnetic field of several hundreds of Gausses. The Magnetic field is used to eliminate the level degeneracy, leading to the frequency-dependent scattering and absorption difference between two perpendicular components of light with right-handed and left-handed circular polarizations. Among the sub-Doppler methods, SAS (Saturated Absorption Spectroscopy) is remarkable. Frequency locking can also be implemented using a stable Fabry-Perot resonator [3-6].

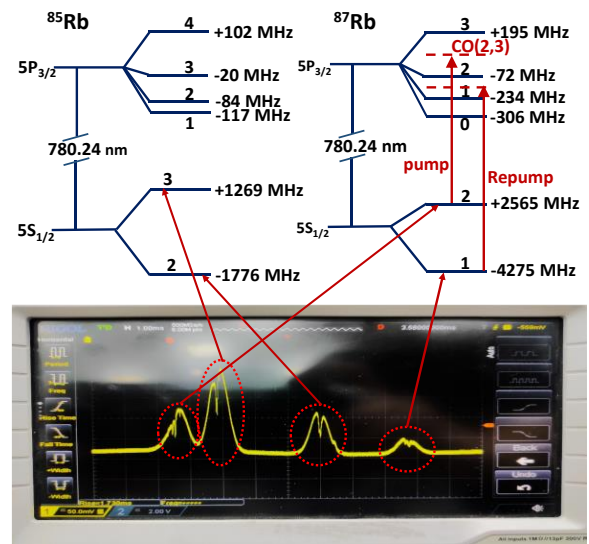

Figure 1. The energy levels and hyperfine structure of two isotopes of rubidium (Rb) atom (D$_2$ Line) and the pump and repump transitions along with the Rb spectrum taken by the oscilloscope in the quantum sensing and metrology lab at ICQT.

In this article, the SAS technique, a prevalent method for locking and stabilizing laser phase fluctuations, is used. A tunable Distributed Brag Reflector (DBR) laser, which can serve as a pump laser in cooling, is employed for frequency locking (as shown in Figure 2). As this figure shows, a part of the DBR laser beam is separated in order to lock and stabilize the frequency. A portion of this beam is separated as a reference beam and detected by APD 1. The remaining part is split into two: a high-intensity pump beam and a low intensity probe beam. These beams pass counterpropagating through the desired gas, and the probe beam is detected by APD 2. The pump beam saturates the atomic transitions, rendering the medium nearly transparent for the probe beam to pass through. It worth mentioning that the Lorentzian line shape profile with a natural line width $\Gamma=\gamma/2\pi$, where $\gamma$ is the probability of spontaneous emission per time, is given by equation (1).

$$\mathscr{L}(\nu,\nu_0) = \frac{1}{1+4(\nu-\nu_0)^2/\Gamma^2} \quad (1)$$

where $\nu$ is the laser frequency and $\nu_0$ is the resonant frequency. When considering Doppler broadening, the Gaussian line shape profile is represented by equation (2).

$$\mathscr{L}(\nu,\nu_0) = \frac{2\sqrt{ln2}}{\sqrt{\pi}\Delta\nu_D} exp\left[-\left(\frac{2\sqrt{ln2}}{\Delta\nu_D}(\nu-\nu_0)\right)^2\right] \quad (2)$$

In equation (2), $\Delta\nu_D = 2\sqrt{\frac{2k_B T ln2}{mc^2}}\nu_0$, is the Doppler linewidth, $k_B$ is Boltzmann's constant, T is the absolute temperature, m is the mass of the particle and c is the speed of light [7].

In this setup, the differential signal between the reference and the probe beams, is sent to the servo as an error signal. The servo amplifies this signal and using a Proportional-Integral-Derivative (PID) controller, generates a control signal which is then sent to the laser controller. For this purpose, the laser frequency is initially modulated externally (Ramp) at a repetition rate of 500 Hz. This modulation allows for a continuous change in the laser frequency, sweeping the absorption frequency region of the rubidium atoms and creating a saturated absorption spectrum.

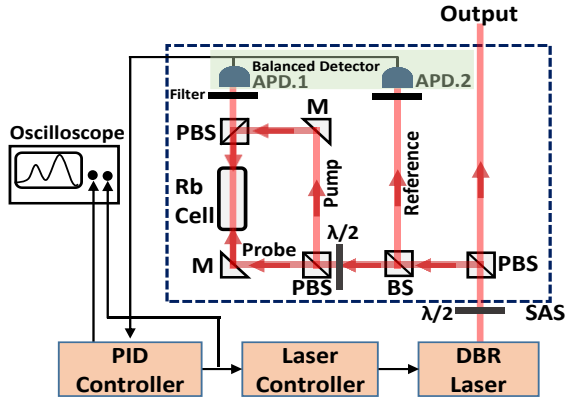

Figure 2. Schematic diagram of DBR laser lock setup using SAS and servo system.

The error signal, which is the difference between the signals from the two detectors and its derivative), is monitored on the oscilloscope. The servo parameters (such as offset) are adjusted so that the error signal becomes zero at the desired frequency, thereby achieving frequency lock. The control signal, sent from the servo to the laser controller, is adjusted in such a way that the error signal in the desired spectral line becomes zero. By this way, when the lock is activated, any unwanted changes in parameters like current, temperature, or environmental factors, that cause a shift in frequency, is balanced by increment or decrement of the laser control voltage. This adjustment continues until the laser frequency returns to the desired frequency.

As we know, in a DBR laser, the laser frequency varies with changes in both temperature and laser current. To facilitate frequency adjustment, the laser temperature is maintained constant with changes of $0.1 m°C/hr$ and the laser frequency is controlled by its current. As mentioned before, by using a SAS system, servo and oscilloscope, the saturated absorption spectrum of rubidium atom and the corresponding error signal are generated. An example of the saturated absorption spectrum of the $D_2$ transition of Rb atoms is shown in Figure 3. Rubidium gas is typically composed of its two naturally occurring isotopes, $^{85}$Rb (72%) and $^{87}$Rb (28%). In the spectrum shown in Figure 3, the peak on the left corresponds to isotope $^{87}$Rb and the peak on the right corresponds to isotope $^{85}$Rb as per Figure 1.

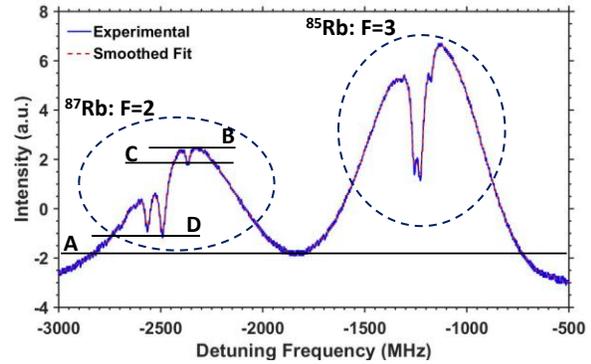

Figure 3: The saturation absorption spectrum of rubidium (Rb) atom was obtained at 39.5 °C using the DBR pump laser in the quantum sensing and metrology lab at ICQT. The specifications of this laser: average laser linewidth, 0.5 MHz, average achievable frequency (without mode hopping), 30 GHz, polarization, horizontal linear and divergence, < 1.3 mrad.

From this spectrum, parameters such as Doppler depth of $^{87}$Rb i.e. ($\frac{A-B}{A}$), hyperfine transition depth of $^{87}$Rb i.e. ($\frac{B-C}{A}$) and the crossover hyperfine transition depth of $^{87}$Rb i.e. ($\frac{D-B}{A}$) can be determined. The values obtained in this experiment and their comparison with the standard values corresponding to the rubidium atom are presented in Table 1. As can be seen, the experimental values align with the standard values of the rubidium atom, thereby validating the obtained spectrum.

Table 1: Values of some spectroscopy parameters of rubidium atom.

| Quantity | Standard Value (%) | Experimental Value (%) |
|---|---|---|
| Doppler Depth | >30 | 236 |
| Hyperfine Doppler Depth | >2.5 | 38.3 |
| Crossover Doppler Depth | >15 | 256 |

Figure 4 illustrates the error signal (depicted in blue) and the servo control voltage (depicted in yellow) as displayed by the oscilloscope in two different scenarios: before and after the frequency locking process. In the locked state, as shown in the figure, the error signal is zero, and the control voltage remains constant, indicating that the desired frequency has been achieved and maintained.

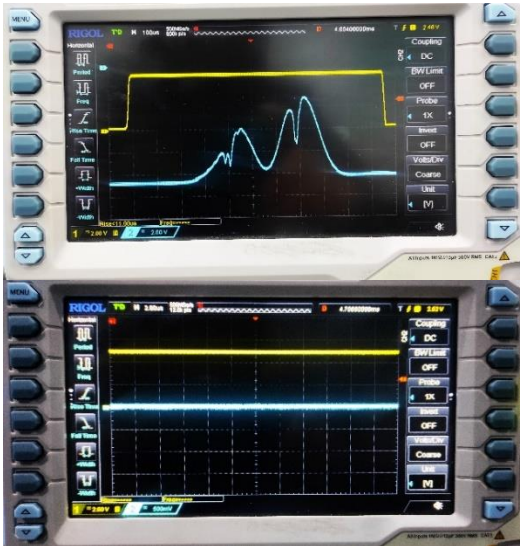

Figure 4. The experimental recorded error signal (depicted in blue) and the servo control voltage (depicted in yellow) in the quantum sensing and metrology lab at ICQT in two different scenarios: before frequency locking (top picture) and after frequency locking (bottom picture) with the DBR laser.

To ensure the stability of the frequency lock, a series of tests were conducted. In the first stage, while in frequency lock mode, it was observed that the error signal remained at zero even when the Ramp was applied, and the spectrum was not visible. In the second test, the light output from the Saturated Absorption Spectroscopy (SAS) was passed through another cell containing rubidium atomic vapor, and its fluorescence radiation was observed. Figure 5 presents the images captured by the CCD, showing the fluorescence radiation of this cell in various states. As depicted in Figure 5-b, when the frequency is locked, the fluorescence emission is clearly visible. However, when the lock is turned off and the system deviates from the transition frequency, the fluorescence radiation diminishes (as shown in Figure 5-c) and eventually extinguishes (as shown in Figure 5-d).

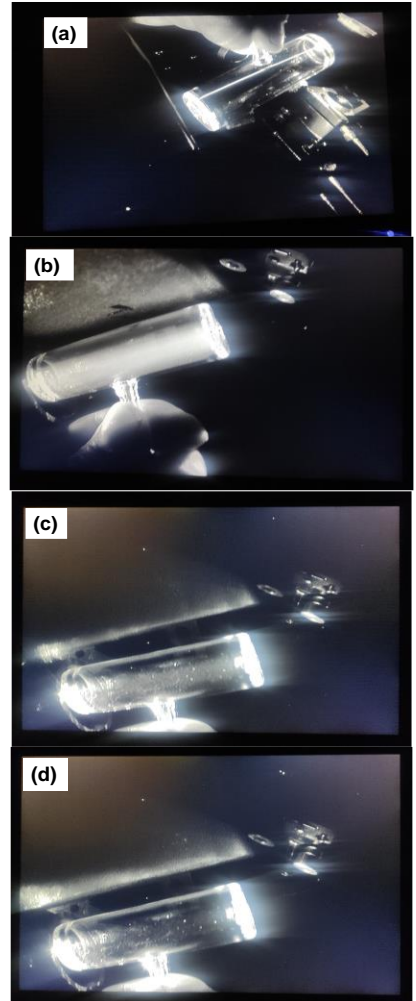

Figure 5. Fluorescence images of the cell containing rubidium atomic vapor in various states of a) frequency lock (laser beam), b) frequency lock (expanded laser beam), c) frequency lock off, low detuning, and d) frequency lock off, large detuning, recorded by the CCD, in the quantum sensing and metrology lab at ICQT.

In an additional test, the voltage transmitted from the servo to the laser controller was measured. In this way, after locking the laser frequency, the laser's temperature was intentionally altered, and the resulting change in control voltage was recorded. A voltage change of 2.8V was observed for a temperature change of 0.1°C. This result demonstrates the servo's effective performance in maintaining a constant laser frequency.

As a perspective, the quantum sensing and metrology group intends to study the effect of magnetic field changes and Zeeman splitting on rubidium atomic levels. Then, with the help of trapping atoms in a magnetic trap, as well as suitable thrust of the laser from opposite directions, the kinetic energy of the atoms is reduced and cooled so that pure quantum effects appear. At that point, it is possible to use the cold atomic source to measure the quantum field, gravity, acceleration, and time.